\documentclass{article}
\usepackage[utf8]{inputenc}
\usepackage[margin=1in]{geometry}

\usepackage{graphicx}      
\usepackage{hyperref}
\usepackage{amsfonts}
\usepackage{amsmath}
\usepackage{amssymb}
\usepackage{bm}
\usepackage{algorithm}
\usepackage{algpseudocode}
\usepackage{lipsum}
\usepackage{color}
\usepackage{hyperref}

\newtheorem{thm}{Theorem}

\def \R {\mathbb{R}}

\def \stablength {T}
\def \nooftasks {m}
\def \noofepochs {k}
\def \FBscale {\sigma_G}
\def \ditherscale {\sigma_{\eta}}
\def \statedim {d_x}
\def \controldim {d_u}
\def \jointtruth {\boldsymbol{\Theta}^{\star}}
\def \noofbases {\ell}
\def \alltruebases {\boldsymbol{\Gamma}^\star}
\def \alltrueweights {\boldsymbol{\omega}^{\star}}

\def \weightgeneric {\boldsymbol{\omega}}
\def \basesgeneric {\boldsymbol{\Gamma}}
\def \Qmatrix {Q}
\def \Rmatrix {R}

\newcommand{\Mnorm}[2]{{\left\vert\kern-0.35ex\left\vert\kern-0.35ex\left\vert #1 \right\vert\kern-0.35ex\right\vert\kern-0.35ex\right\vert}}
\newcommand{\norm}[2]{{\left\vert\kern-0.35ex\left\vert #1 \right\vert\kern-0.35ex\right\vert}}

\newcommand{\normaldist}[2]{\mathcal{N}\left(#1,#2\right)}
\newcommand{\Gainmatrix}[1]{G_{#1}}
\newcommand{\OptGain}[1]{G\left(#1\right)}
\newcommand{\RiccatiSolution}[1]{P\left(#1\right)}
\newcommand{\noise}[2]{\xi_{#1}\left(#2\right)}
\newcommand{\dither}[2]{\eta_{#1}\left(#2\right)}
\newcommand{\state}[2]{x_{#1}\left(#2\right)}
\newcommand{\control}[2]{u_{#1}\left(#2\right)}
\newcommand{\regressors}[2]{z_{#1}\left(#2\right)}

\newcommand{\truth}[1]{{\theta}^{\star}_{#1}}
\newcommand{\truebasismat}[1]{{\Gamma}^{\star}_{#1}}
\newcommand{\basismat}[1]{{\Gamma}_{#1}}
\newcommand{\trueweights}[2]{\omega^{\star}_{#1,#2}}
\newcommand{\weights}[2]{\omega_{#1,#2}}
\newcommand{\trueA}[1]{{A}^{\star}_{#1}}
\newcommand{\trueB}[1]{{B}^{\star}_{#1}}
\newcommand{\jointestimate}[1]{\boldsymbol{\widehat{\Theta}_{#1}}}
\newcommand{\indivestimate}[2]{\widehat{\theta}_{#1,#2}}

\newcommand{\epochtimes}[1]{\tau_{#1}}
\newcommand{\eigmax}[1]{\rho \left(#1\right)}

\title{\bf Joint Learning-Based Stabilization of Multiple Unknown Linear Systems} 

\author{Mohamad Kazem Shirani Faradonbeh, Aditya Modi}
\date{}

\begin{document}
	
	\maketitle

	\begin{abstract}
		Learning-based control of linear systems received a lot of attentions recently. In popular settings, the true dynamical models are unknown to the decision-maker and need to be interactively learned by applying control inputs to the systems. Unlike the matured literature of efficient reinforcement learning policies for adaptive control of a single system, results on joint learning of multiple systems are not currently available. Especially, the important problem of fast and reliable joint-stabilization remains unaddressed and so is the focus of this work. We propose a novel joint learning-based stabilization algorithm for quickly learning stabilizing policies for all systems understudy, from  the data of unstable state trajectories. The presented procedure is shown to be notably effective such that it stabilizes the family of dynamical systems in an extremely short time period.
	\end{abstract}

	\section{Introduction}
	
	Study of reinforcement learning algorithms for sequential learning-based decision-making in unknown linear systems has become increasingly popular in the recent years. In the canonical version of the problem, the true dynamics matrices of the plant are unknown, and the goal consists of adaptive design of the control input for minimizing deviations from optimal policy. Still, the control actions must be diverse enough to lead to accurate identification of the unknown parameters~\cite{lai1985asymptotic}. The existing literature is notably rich, including adaptive policies based on optimistic approximations of the dynamics matrices over a confidence region~\cite{abbasi2011regret,faradonbeh2020optimism}, as well as plugin estimates of unknown parameters after leveraging a dither signal~\cite{faradonbeh2020input,lale2020explore,ziemann2021applications}, Bayesian approaches~\cite{abeille2018improved,faradonbeh2020adaptive,sudhakara2021scalable}, and statistical bootstrap~\cite{faradonbeh2019applications}.
	
	An important problem in in different areas of control theory is that of stabilization. In adaptive control that the true dynamical model is not available, a reinforcement learning algorithm needs to stabilize the system, prior to designing and applying an efficient policy for minimizing the cost function. For a single system, stabilization procedures based on confidence regions around dynamics matrices~\cite{abbasi2011regret,faradonbeh2018bfinite,lale2020explore} and plugin estimates of them~\cite{shirani2017non,faradonbeh2019randomized,chen2021black} are proposed. 
	
	However, an interesting version of the problem that aims to stabilize multiple unknown linear systems is not addressed in the existing literature and so will be considered in this work. It is shown for the simpler version of the problem that by utilizing \emph{known} similarities among a group of sub-systems, the total regret of reinforcement learning policies can decrease~\cite{sudhakara2021scalable}. However, the more interesting version of the problem that similarities are not known is not addressed. Indeed, in the existing literature it is assumed that stabilization of the systems is a priori completed and stabilizing policies are provided to the reinforcement learning policy~\cite{du2019continuous}. 
	
	In this work, we study joint stabilization of multiple unknown stochastic linear dynamical systems in a general framework and under minimal assumptions. That is, we consider a joint estimator that can learn dynamics matrices of multiple linear systems, without having access to the similarities among them. Technically, the presented procedure jointly learns the linear systems in a commonly used and realistic setting that the true dynamics matrices are \emph{unknown} linear combinations of some \emph{unknown} basis matrices (\cite{maurer2016benefit}, see the equation in~\eqref{SharedBasesAssumpEq}). 
	
	For learning multiple linear systems, a recent work establishes that if all systems are (marginally) stable, the identification error rates significantly decrease by jointly learning the dynamics matrices~\cite{modi2021joint}. Therefore, notably short state-input trajectories are capable of providing accurate estimates. However, since during a stabilization procedure the system is with high probability unstable~\cite{lai1985asymptotic}, the existing analysis is not applicable to the setting of this work. Accordingly, we aim to study if joint-stabilization can guarantee system stabilities faster. Note that for stabilization purposes, shortness of trajectories is extremely crucial since unstable evolution of the plants for a long time, defeats the purpose and is exactly what we aim to preclude.
	
	The structure of this work is provided next. In Section~\ref{SetupSection}, we discuss the backgrounds and relevant technical preliminaries. That includes the precise formulation of the problem, followed by brief review of stabilization of unknown linear systems, as well as an overview of joint-learning of multiple unknown systems. Then, in Section~\ref{AlgoSection}, we introduce an algorithm for joint learning-based stabilization and discuss its details together with intuitions of different steps. The analysis of the proposed algorithm as well as interpretation of the results and tuning of parameters presented in Section~\ref{NumericalSection}. Finally, we investigate future directions for extending the results and express interesting problems that this work paves the roads towards.
	
	\paragraph*{Notation:} For a positive integer $n$, define $[n]=\left\{1, \cdots, n\right\}$. For vectors, we use $\norm{\cdot}{2}$ to refer to Euclidean norm. To measure matrices, we use the operator norms $\Mnorm{M}{}$, which for $M \in \R^{d_1 \times d_2}$, is defined as the maximum value of $\norm{Mv}{2}/\norm{v}{2}$ among all non-zero $v \in \R^{d_2}$. Moreover, $\eigmax{M}$ denotes the spectral radius of the square matrix $M$. For a linear dynamical control system that with the open-loop transition matrix $A$ and the input matrix $B$, we use $\theta=\left[A,B\right]^\top$ to denote the dynamics parameter (see~\eqref{dynamics}).
	
	\section{Setup and Preliminaries } \label{SetupSection}
	In this section, we discuss the problem of joint stabilization for multiple unknown linear system. To that end, we first introduce the technical framework including a general setting that relates the dynamics parameters of different systems. Then, we discuss the widely-used concept of Riccati equation, followed by its applications as a universal standard method for stabilizing linear systems based on \emph{fully accurate} knowledge of the dynamics matrices. 
	
	Next, the statistical learning framework for jointly estimating multiple linear systems will be briefly explained. We present a reliable estimator that utilizes possible similarities among the systems understudy for accurate identification of the unknown dynamics parameters. The proposed method fully learns the dynamics matrices from the observed state trajectories such that \emph{no knowledge} about the similarities between the systems is required.
	
	\subsection{Problem Formulation}
	Suppose that there are $\nooftasks$ linear dynamical systems, where system $i \in [\nooftasks]$ follows the stochastic state equation
	\begin{eqnarray} \label{dynamics}
	\state{i}{t+1}=\trueA{i} \state{i}{t} + \trueB{i} \control{i}{t} + \noise{i}{t}.
	\end{eqnarray}
	Above, the state of system $i$ at time $t$ is represented by $\state{i}{t} \in \R^{\statedim}$, while $\control{i}{t} \in \R^{\controldim}$ is the control action applied to system $i$ at time $t$ and $\noise{i}{t}$ represent the exogenous stochastic disturbance. Therefore, a control policy designs $\control{i}{t}$ for all $i \in [\nooftasks], t \geq 0$ and at time $t$ applies $\control{i}{t}$ to system $i$. Then, the policy observes the resulting state trajectories $\state{i}{t}, i \in [\nooftasks]$ over time. Note that the decision-maker is completely oblivious about the disturbance noise. The overall goal is to ensure that all systems perform in a stable manner, as elaborated below.
	
	Importantly, all dynamics matrices $\truth{i}=\left[\trueA{i},\trueB{i}\right]^\top$ are unknown to the control policy. Thus, design of the control inputs for all systems is merely based on the observed data of multidimensional state trajectories. More precisely, a reliable method for designing the input and using the observations for accurately learning $\left\{\trueA{i},\trueB{i}\right\}_{i=1}^\nooftasks$ is required for learning to stabilize the systems. Note that for stabilization, approximations of dynamics parameters are required for all $\nooftasks$ dynamical systems. Furthermore, the lengths of the trajectories is extremely limited since uncertainties about the dynamical models in~\eqref{dynamics} destabilize (a majority or all of) the systems. Accordingly, one needs to learn under \emph{data scarcity} since unstable dynamics cannot operate in long-term. This issue of limited data renders joint learning doubly important. So, the goal is to learn to stabilize fast and under instabilities, as well as to study effectiveness of doing so in a jointly manner. 
	
	To proceed, denote the true dynamics parameter of system $i$ by $\truth{i}=\left[\trueA{i},\trueB{i}\right]^{\top} \in \R^{(\statedim+\controldim) \times \statedim}$. To have a nontrivial problem, we assume that all systems are stabilizable in the sense that for all $i \in [\nooftasks]$, there exists some unknown  $\Gainmatrix{i} \in \R^{\controldim \times \statedim}$ such that all eigenvalues of the closed-loop matrices $\trueA{i}+\trueB{i}\Gainmatrix{i}$ are inside the unit-circle (i.e., their magnitudes as complex numbers are less than $1$). Clearly, this is the most general setting that stabilization of linear dynamical systems is feasible.
	
	\subsection{Stabilization of Linear Systems} \label{StabSubsection}
	Now, we discuss the hypothetical situation of stabilizing linear systems with exactly known dynamics matrices and Riccati equations. It is well-known that stabilizability is a sufficient and necessary condition for existence of solutions for Riccati equations for strictly convex quadratic cost functions~\cite{chan1984convergence,de1986riccati,faradonbeh2018bfinite}. Furthermore, Riccati equations provide a stabilizing feedback in a computationally fast procedure. To proceed, let $\Qmatrix \in \R^{\statedim \times \statedim}$ and $ \Rmatrix \in \R^{\controldim \times \controldim}$ be arbitrary positive definite matrices. Then, suppose that the positive semidefinite matrix $\RiccatiSolution{\truth{i}}$ solves the fixed-point matrix equation 
	\begin{equation} \label{RiccatiEq}
	\RiccatiSolution{\theta} = \Qmatrix + A^\top \RiccatiSolution{\theta} A - A^\top \RiccatiSolution{\theta} B \left( B^\top \RiccatiSolution{\theta} B + \Rmatrix \right)^{-1} B^\top \RiccatiSolution{\theta} A,
	\end{equation}
	for $A=\trueA{i}, B=\trueB{i}$, and use $\RiccatiSolution{\truth{i}}$ to define the feedback matrix 
	\\$\OptGain{\truth{i}} = - \left( {\trueB{i}}^\top \RiccatiSolution{\truth{i}} \trueB{i} + \Rmatrix \right)^{-1} {\trueB{i}}^\top \RiccatiSolution{\truth{i}} \trueA{i}$.
	
	So, the algebraic Riccati equation in \eqref{RiccatiEq} provides stabilizing feedback matrices $\OptGain{\truth{i}}$, for all systems $i =1, \cdots, \nooftasks$, as stated below. 
	\begin{thm} \label{KnownStabThm}
		It holds that $ \eigmax{\trueA{i}+\trueB{i}\OptGain{\truth{i}}}<1$.
	\end{thm}
	Note that $\truth{i}$ is unknown and so $\OptGain{\truth{i}}$ is unavailable. However, it is shown that coarse-grained approximations of $\truth{i}, i \in [\nooftasks]$ are sufficient for stabilizing the systems such that if $\Mnorm{\indivestimate{i}{}-\truth{i}}{arg2}$ is small enough, $\OptGain{\indivestimate{i}{}}$ suffices for stabilization and we have $\eigmax{\trueA{i}+\trueB{i}\OptGain{\indivestimate{i}{}}}<1$. Technical statements and details are beyond the scope of this work and can be found in the relevant literature~\cite{shirani2017non,faradonbeh2018bfinite,faradonbeh2019randomized,lale2020explore,chen2021black}.

	\subsection{Joint Learning of Linear Systems}
	Next, we discuss joint learning of multiple unknown linear dynamical systems. Later on, we propose an algorithm for utilizing the following joint-learning procedures for learning-based stabilization of the unknown systems understudy. First, we rewrite the dynamics in \eqref{dynamics} based the aforementioned notation $\truth{i}$, together with the observation vector of system $i$ at time $t$, denoted by $\regressors{i}{t}=\left[\state{i}{t}^\top,\control{i}{t}^\top\right]^\top$. So, the data generation model is
	\begin{equation} \label{RegressionEq}
	\state{i}{t+1} = {\truth{i}}^\top \regressors{i}{t} + \noise{i}{t}, i \in [\nooftasks], 0 \leq t \leq \stablength,
	\end{equation}
	for some joint-stabilization time length $\stablength$. Clearly, it is of crucial importance to find the smallest possible $\stablength$, yet ensuring that stabilization will be successful.
	
	We assume that the systems understudy share some commonalities such that for some shared bases $\truebasismat{j} \in \R^{(\statedim+\controldim) \times \statedim}, j \in [\noofbases]$, it holds that
	\begin{equation} \label{SharedBasesAssumpEq}
	\truth{i} = \sum\limits_{j=1}^{\noofbases} \trueweights{i}{j} \truebasismat{j},
	\end{equation}
	where $\trueweights{i}{j}$ are the coefficients reflecting the contributions of shared bases in constructing the dynamics matrices of the systems. Importantly, \emph{all} the coefficients $\trueweights{i}{j}$ and matrices $\truebasismat{j}$ are fully \emph{unknown} and need to be learned from the data. Settings of the form of \eqref{SharedBasesAssumpEq} are ubiquitous in the literature of jointly learning multiple parameters and provide a realistic and effective framework for utilizing (possibly unknown) similarities among quantities of interest in different disciplines~\cite{maurer2006bounds,maurer2016benefit,du2020few,tripuraneni2020provable}.
	
	For succinctness, denote $\jointtruth=\left[\truth{1}, \cdots, \truth{\nooftasks}\right]$, $\alltruebases=\left[\truebasismat{1}, \cdots, \truebasismat{\noofbases}\right]$, and $\alltrueweights=\left[ \trueweights{i}{j} \right]_{1 \leq i \leq \nooftasks, 1 \leq j \leq \noofbases}$. So, the goal of learning is to approximate $\jointtruth$ leveraging the fact that it accepts the representation in \eqref{SharedBasesAssumpEq} based on $\alltruebases,\alltrueweights$. For this purpose, \eqref{RegressionEq} recommends that we compute $\jointestimate{\stablength}$ that minimizes the total sum of squares of deviations $\state{i}{t+1} - {\truth{i}}^\top \regressors{i}{t}$;
	\begin{equation} \label{JointLearningStepEq}
	\jointestimate{\stablength} = \arg\!\min\limits_{\basesgeneric, \weightgeneric} \sum\limits_{i=1}^\nooftasks \sum\limits_{j=1}^{\noofepochs} \frac{1}{ \norm{\regressors{i}{\epochtimes{j-1}}}{2}^2 \vee 1} \sum\limits_{t=\epochtimes{j-1}}^{\epochtimes{j}-1} \norm{\state{i}{t+1} - \left( \sum\limits_{j=1}^{\noofbases} \weights{i}{j} \basismat{j} \right)^\top \regressors{i}{t}}{arg2}^2.
	\end{equation}
	
	Further details of this estimation procedure will be discussed shortly.  Theoretical performance analyses of this estimator for stable systems is recently established~\cite{modi2021joint}. However, effectiveness of such learning methods in unstable dynamics and for learning-to-stabilize is not studied to the date, and is the focus of the subsequent sections. Finally, although the objective quantity of interest is a non-convex function of $\alltruebases,\alltrueweights$, it admits computationally efficient solutions and accurate minimizers can be found fast 
	\cite{bhojanapalli2016global,ge2017no,jain2017non,tripuraneni2020provable}.   
	\begin{table*}
		
	\end{table*}

	
	

	\section{Joint Learning-based Stabilization} \label{AlgoSection}
	Broadly speaking, the joint learning-based stabilization Algorithm~\ref{algo} applies random control inputs in different epochs, collects the observed data, and calculates the joint estimate of all dynamics matrices using a properly defined loss function. Then, Algorithm~\ref{algo} uses learned-parameters to design stabilizing feedbacks for all systems according to discrete time algebraic Riccati equations. The randomization of the control inputs consists of both random feedback matrices as well as additive stochastic dither signals, as elaborated below. 
	
	The inputs of Algorithm~\ref{algo} are the time length $\stablength$ that the algorithm can interact with all dynamical systems, the standard deviation $\FBscale$ that determines the normal distribution from which the columns of random feedback matrices are sampled, the standard deviation of the stochastic dither signals $\dither{i}{t}$ denoted by $\ditherscale$, and the integer $\noofepochs$ that indicates the number of epochs of interacting with the system, such that at the beginning of each epoch a new random feedback matrix is employed.    
	
	Algorithm~\ref{algo} takes $\stablength$ and divides the time window $0 \leq t \leq \stablength$ to $\noofepochs$ epochs of (almost) equal lengths.  During each epoch $j \in [\noofbases]$, a random feedback matrix $\Gainmatrix{j}$ is generated by independently sampling all $\controldim$ columns of $\Gainmatrix{j}$ from the multivariate Gaussian distribution $\normaldist{0}{\FBscale I_{\statedim}}$, where the standard deviation $\FBscale$ is provided to the algorithm as an input parameter. Random feedback matrices of this form are used before for stabilization of a single system~\cite{faradonbeh2019randomized}. The rationale of employing random matrices is that some singularities can occur for a measure-zero set of closed-loop matrices, making acceptable learning impossible~\cite{faradonbeh2018finite}. To avoid these singularities that prevent accurate learning of dynamical systems, the loops in all systems are closed through random linear feedback matrices~\cite{faradonbeh2018bfinite}. Further details are beyond the framework of this paper and can be found in the cited literature.
	
	\begin{algorithm}
		\caption{{\bf: Joint Learning-based Stabilization} } \label{algo}
		\begin{algorithmic}
			\State Inputs: joint-stabilization length $\stablength$, feedback scale $\FBscale$, dither signal scale $\ditherscale$, number of epochs $\noofepochs$ 
			\State Output: stabilizing feedbacks  $\left\{ \OptGain{\indivestimate{i}{T}} \right\}_{i=1}^{\nooftasks}$\\
			\State Let $\epochtimes{0}=0$
			\For{epoch $j=1,\cdots,\noofepochs$}
			\State Draw independent columns of $\Gainmatrix{j}$ from $\normaldist{0}{\FBscale I_{\statedim}}$
			\State Let $\epochtimes{j}=\lfloor j \stablength /\noofepochs \rfloor$
			\For{epoch $\epochtimes{j-1} \leq t < \epochtimes{j}$}
			\For{system $i=1,\cdots,\nooftasks$}
			\State Draw $\dither{i}{t}$ from $\normaldist{0}{\ditherscale I_{\controldim}}$, independently
			\State Apply control action $\control{i}{t} = \Gainmatrix{j} \state{i}{t}+ \dither{i}{t}$
			\EndFor
			\EndFor
			\EndFor
			\State Using \eqref{JointLearningStepEq}, calculate $\jointestimate{\stablength}=\left[ \indivestimate{1}{\stablength}, \cdots, \indivestimate{\nooftasks}{\stablength}\right]$
			\State Return $\left\{ \OptGain{\indivestimate{i}{T}} \right\}_{i=1}^{\nooftasks}$, according to \eqref{OptimalFBMatrixEq} 
		\end{algorithmic}
	\end{algorithm}
	
	The control inputs are also randomized through an additive dither signal, in order to improve exploration, as will be explained shortly. In fact, during epoch $j$, the algorithm employs the control action $\control{i}{t} = \Gainmatrix{j} \state{i}{t}+ \dither{i}{t}$ to system $i$, where the random signals $\dither{i}{t}$ are generated independently from the $\controldim$-dimensional Gaussian distribution $\normaldist{0}{\ditherscale I_{\controldim}}$. Intuitively, adding $\dither{i}{t}$ to the feedback control signal helps to enrich the data and obtain acceptable approximations of the unknown parameters $\alltruebases,\alltrueweights$. In fact, since the covariates $\state{i}{t}$ and $\control{i}{t}$ are highly correlated through the feedback matrix $\Gainmatrix{j}$, the dither signals $\dither{i}{t}$ weaken the correlation rendering the data of state trajectories more diverse and richer. 
	
	Then, the designed control inputs as well as the resulting state vectors of all systems are gathered and used for jointly learning all dynamics parameters. So, the data for designing $\nooftasks$ stabilizing policies consists of $\left\{ \regressors{i}{t} : 1 \leq i \leq \nooftasks, 0 \leq t \leq \stablength \right\}$, as defined before~\eqref{RegressionEq}. Now, the unknown dynamics parameters of all systems, which is denoted by $\jointtruth$, is estimated by \eqref{JointLearningStepEq}. It is based on a sum-of-squares loss function, together with an appropriate rescaling factor so that the estimator will be able to address the (possibly) unstable behavior of the systems which are destabilized by the random feedback matrices $\Gainmatrix{j}$. Thus, the algorithm jointly learns the unknown true dynamics of the systems by calculating $\jointestimate{\stablength}$ in \eqref{JointLearningStepEq}.
	
	Note that in \eqref{JointLearningStepEq}, the time frame $\epochtimes{j-1} \leq t < \epochtimes{j}$ corresponds to epoch $j$. So, the data observed during epoch $j$ forms the deviation $\norm{\state{i}{t+1} - \left( \sum\limits_{j=1}^{\noofbases} \weights{i}{j} \basismat{j} \right)^\top \regressors{i}{t}}{arg2}^2$. Then, adding up over all times $t$ of epoch $j$, we divide the obtained loss of system $i$ by the quantity $\norm{\regressors{i}{\epochtimes{j-1}}}{2}^2$, where $\epochtimes{j-1}$ is the start time of epoch $j$. Then, the summation of the resulting rescaled squared deviations over all systems and all epochs constitutes the total loss function. The reason for adding the above rescaling is that instabilities of the dynamical systems make the state vectors exponentially growing as time grows~\cite{faradonbeh2018finite}. Therefore, state vectors of future epochs are much larger than those of the current epoch. To ensure that data of all epochs are balanced and will be effectively used in the learning procedure, we employ this rescaling factor.
	
	Now, because the joint estimate $\jointestimate{\stablength}=\left[ \indivestimate{1}{\stablength}, \cdots, \indivestimate{\nooftasks}{\stablength}\right]$ contains estimates for all system dynamics, $\indivestimate{i}{\stablength}$ is used to stabilize system $i$. That is, since we expect $\indivestimate{i}{T}$ to provide an acceptable approximation of $\truth{i}$, according to the discussions in Subsection~\ref{StabSubsection}, we expect $\OptGain{\indivestimate{i}{\stablength}}$ to stabilize the system $i$, where
	\begin{equation} \label{OptimalFBMatrixEq}
	\OptGain{\theta} = - \left( {B}^\top \RiccatiSolution{\theta} B + \Rmatrix \right)^{-1} {B}^\top \RiccatiSolution{\theta} A.
	\end{equation} 
	
	Therefore, Algorithm \ref{algo} returns the feedback matrices $\left\{ \OptGain{\indivestimate{i}{T}} \right\}_{i=1}^{\nooftasks}$ as stabilizers for the systems.
	
	\begin{figure}
		\centering
		\resizebox*{14.5cm}{!}{\includegraphics{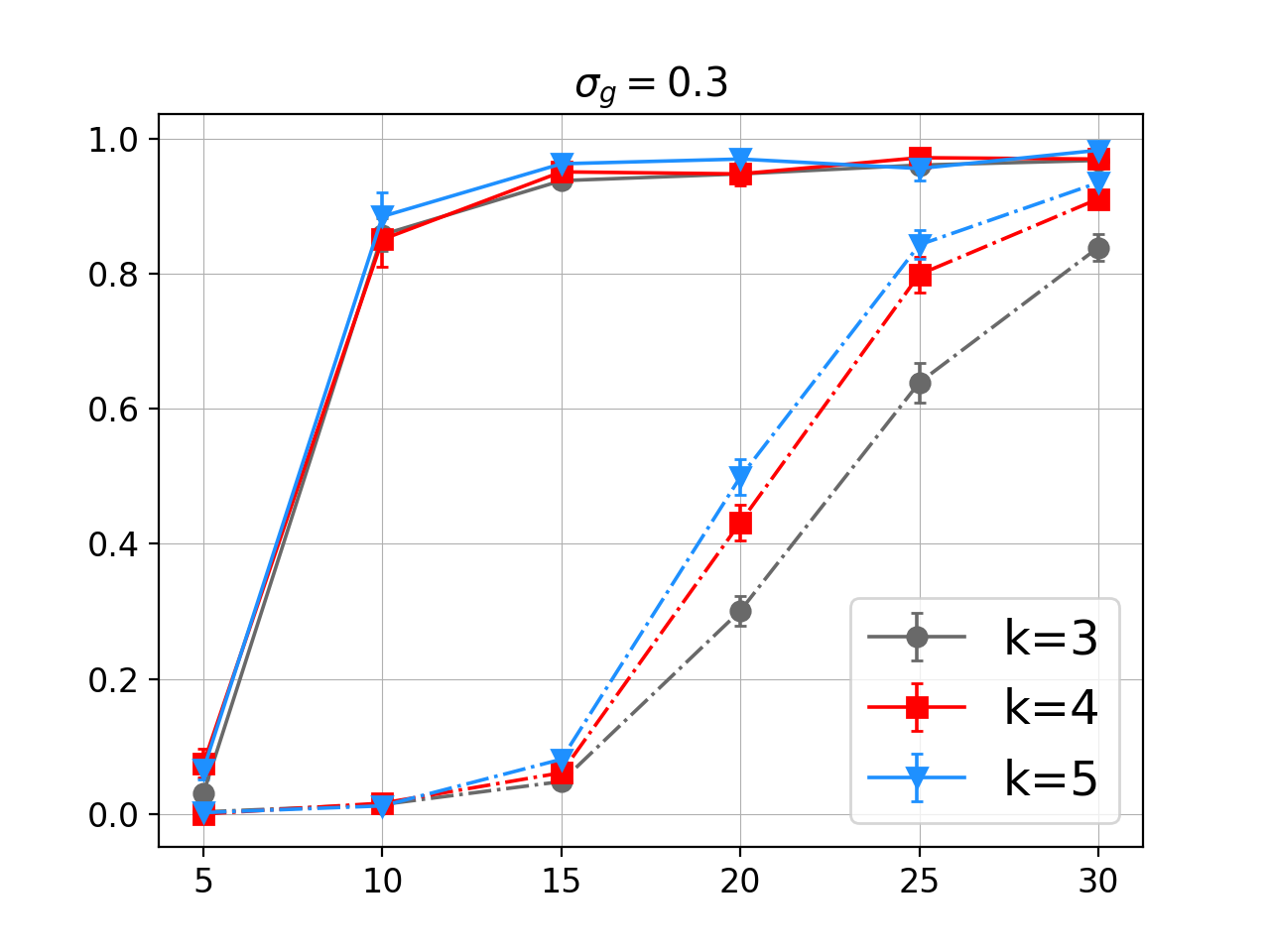}\includegraphics{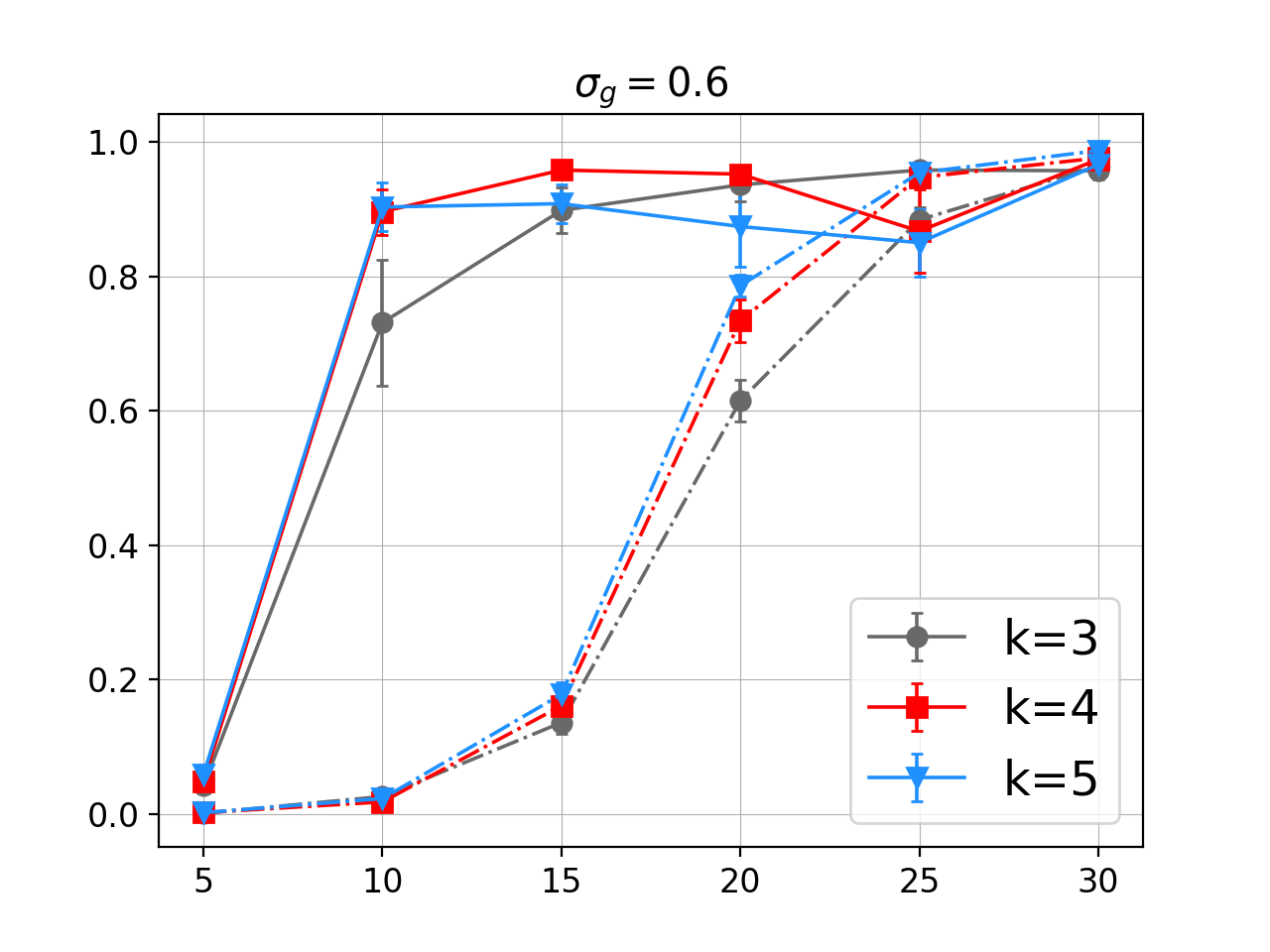}} \resizebox*{14.5cm}{!}{\includegraphics{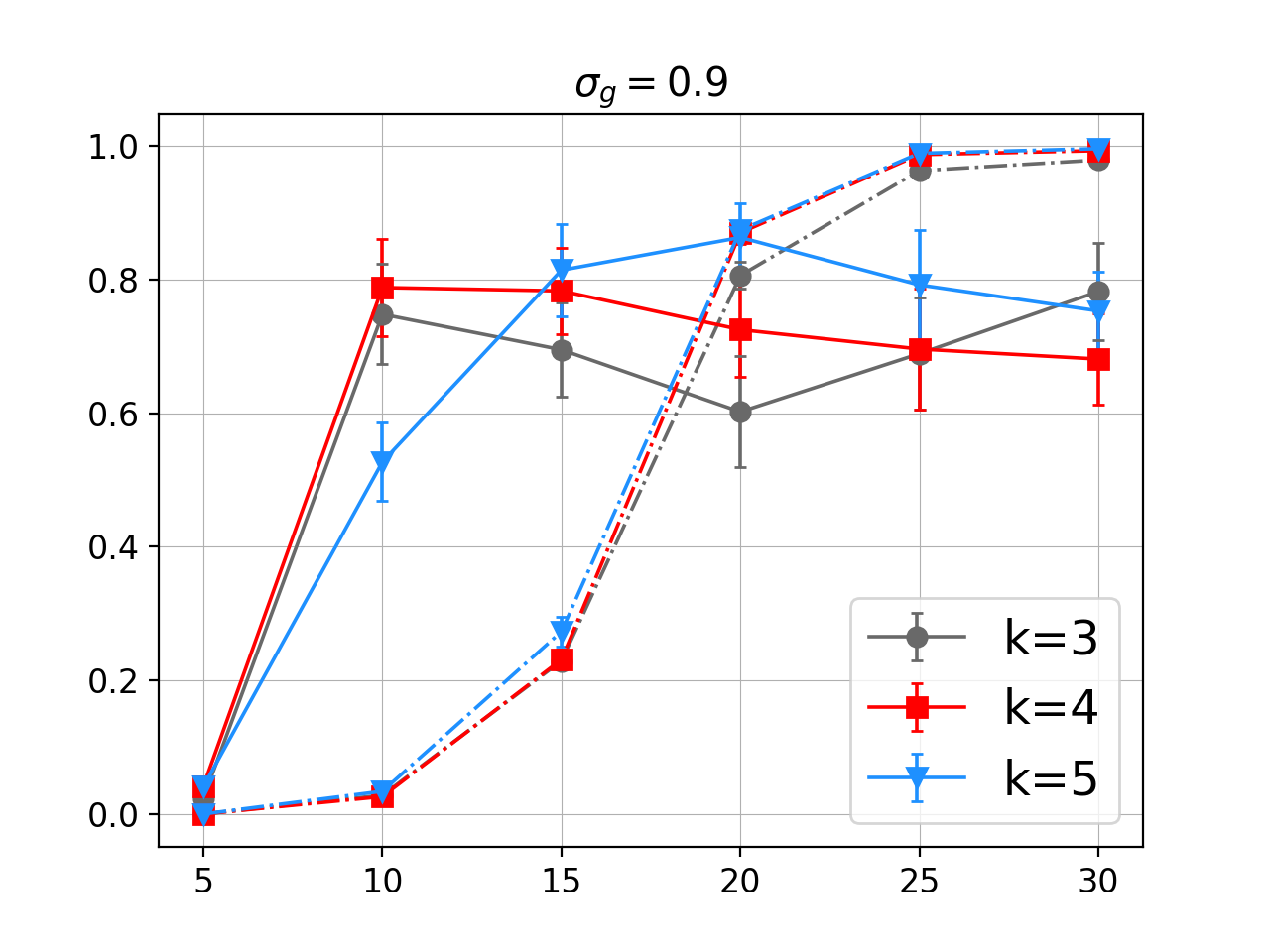}\includegraphics{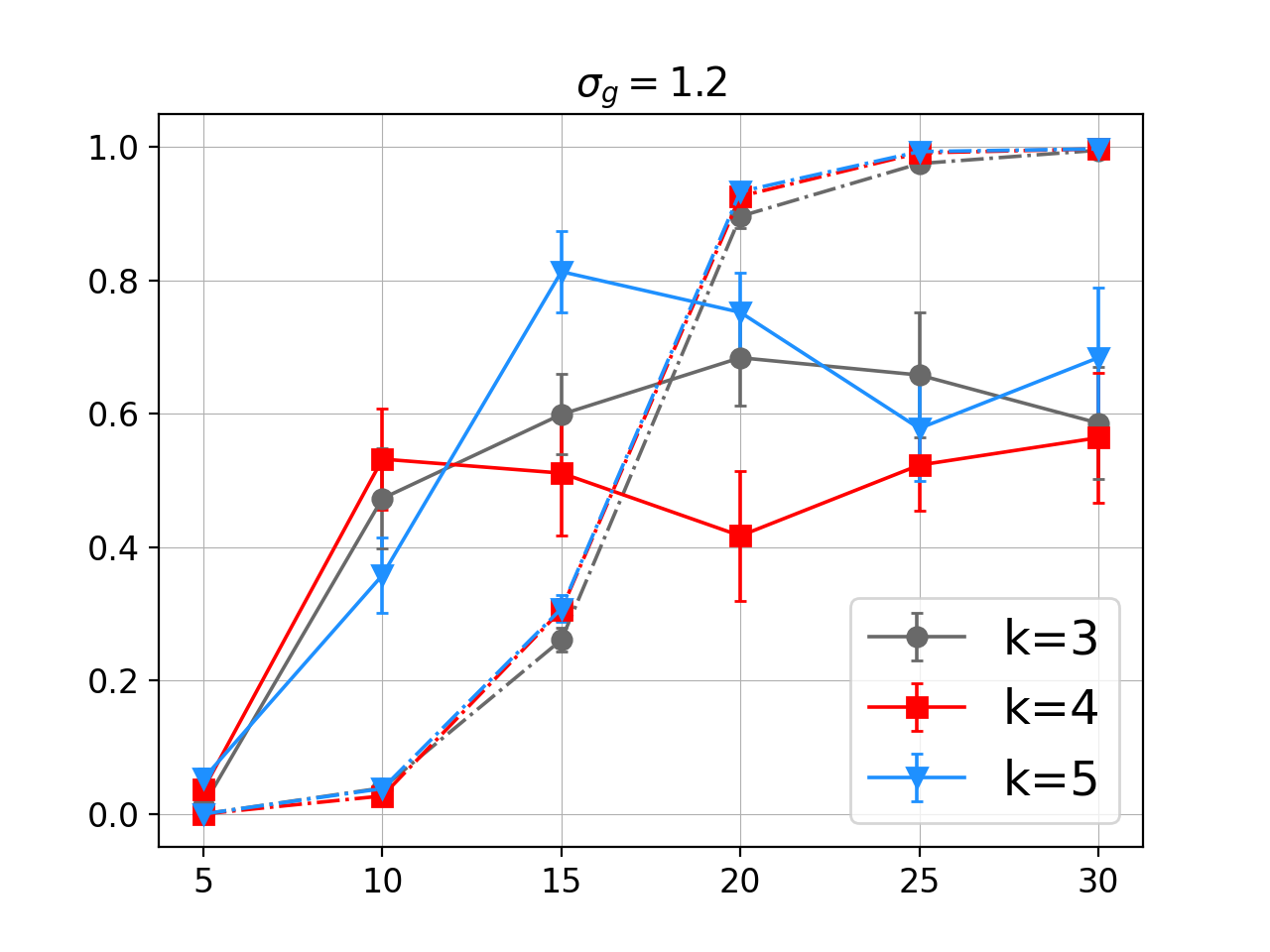}} 
		\caption{The graphs indicate the portion of systems stabilized by Algorithm~\ref{algo} versus the joint-stabilization length $\stablength$. Each plot corresponds to a value $\FBscale \in \left\{ 0.3, 0.6, 0.9, 1.2 \right\}$, for different number of epochs $\noofepochs=3,\textcolor{red}{4},\textcolor{blue}{5}$. Dashed curves present the stabilized portion by individual stabilization methods.}
		\label{fig:sigmaG}
	\end{figure}
	
	\begin{figure}
		\centering
		\resizebox*{14.5cm}{!}{\includegraphics{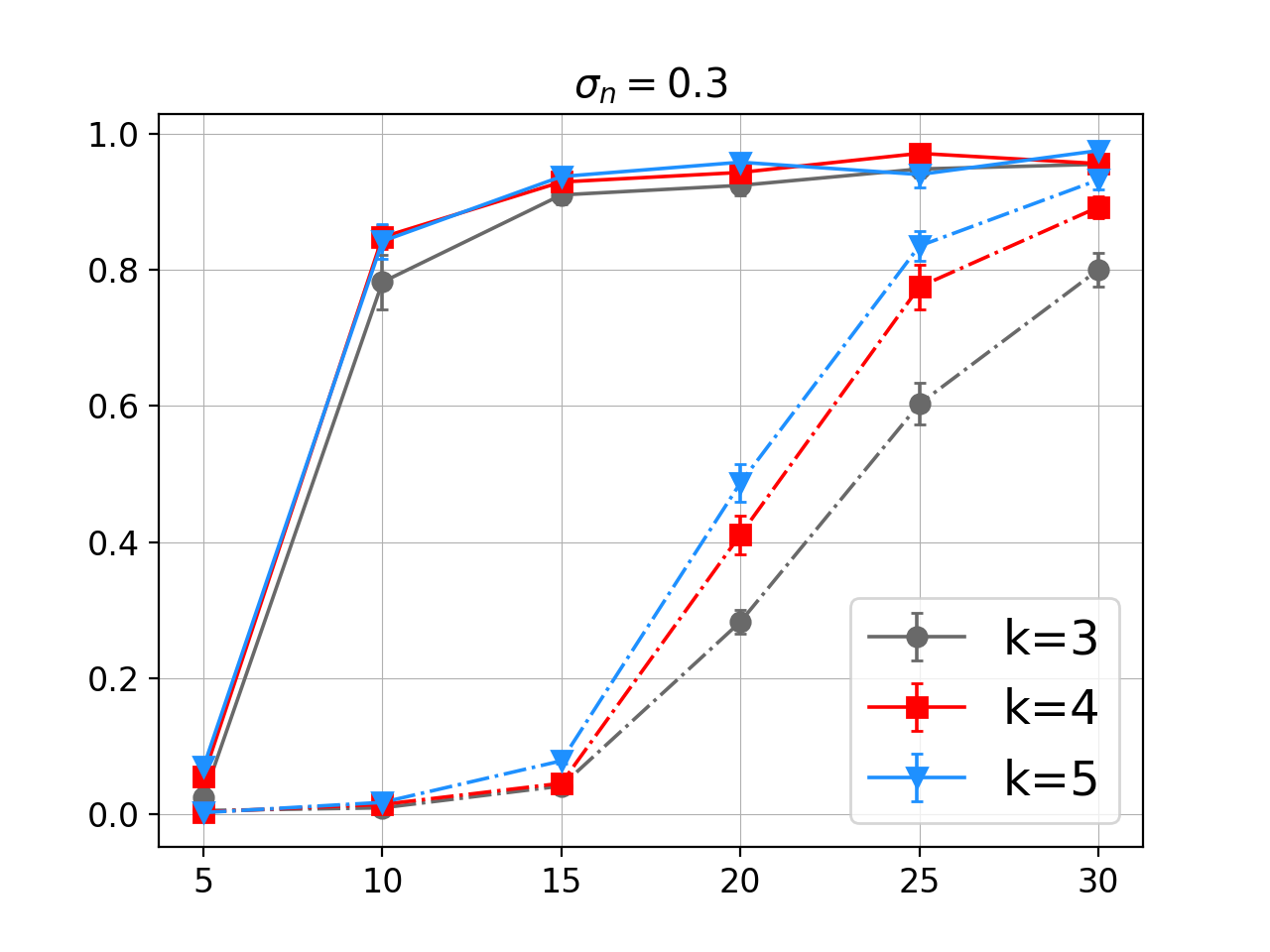}\includegraphics{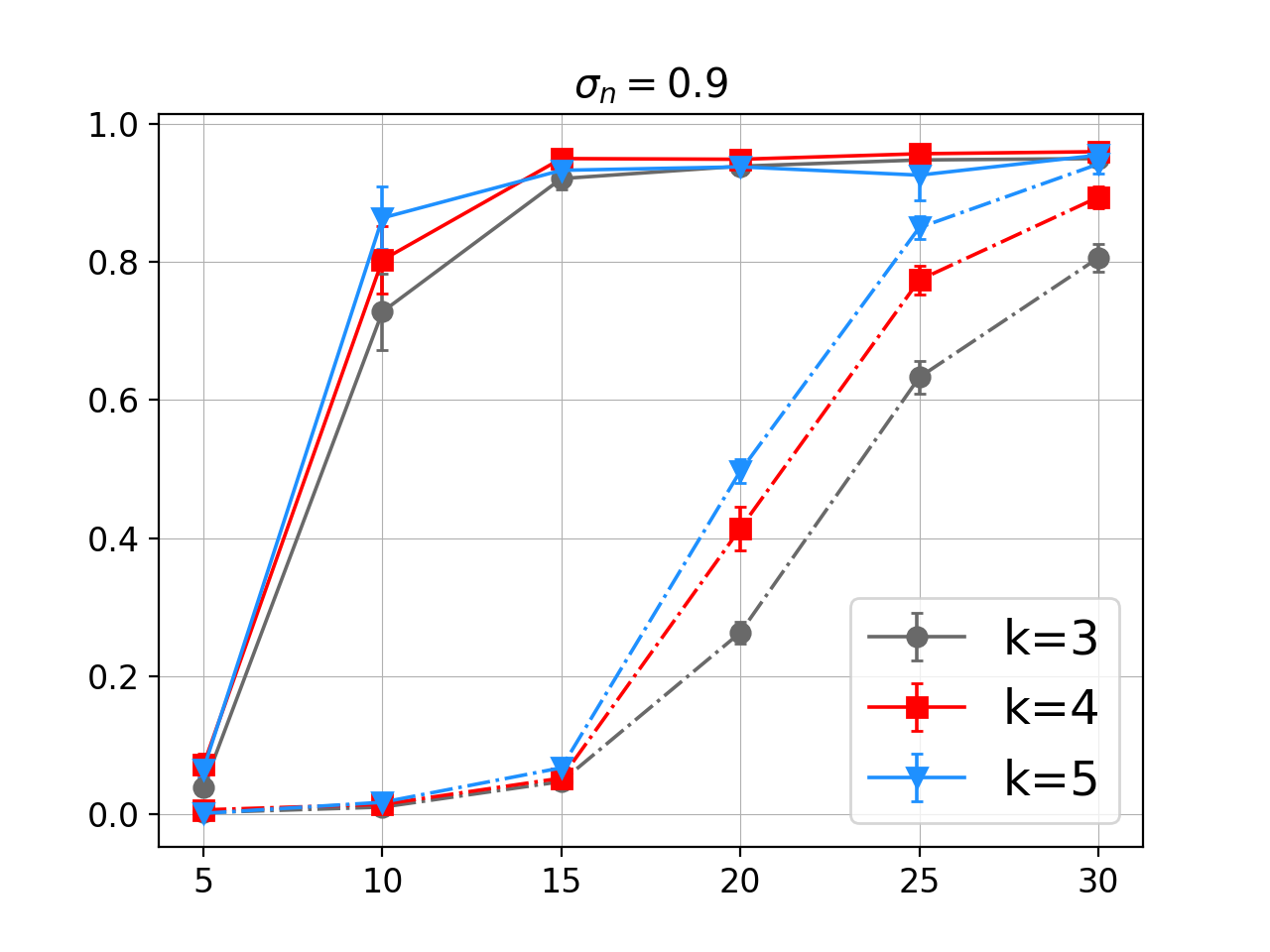}} \resizebox*{14.5cm}{!}{\includegraphics{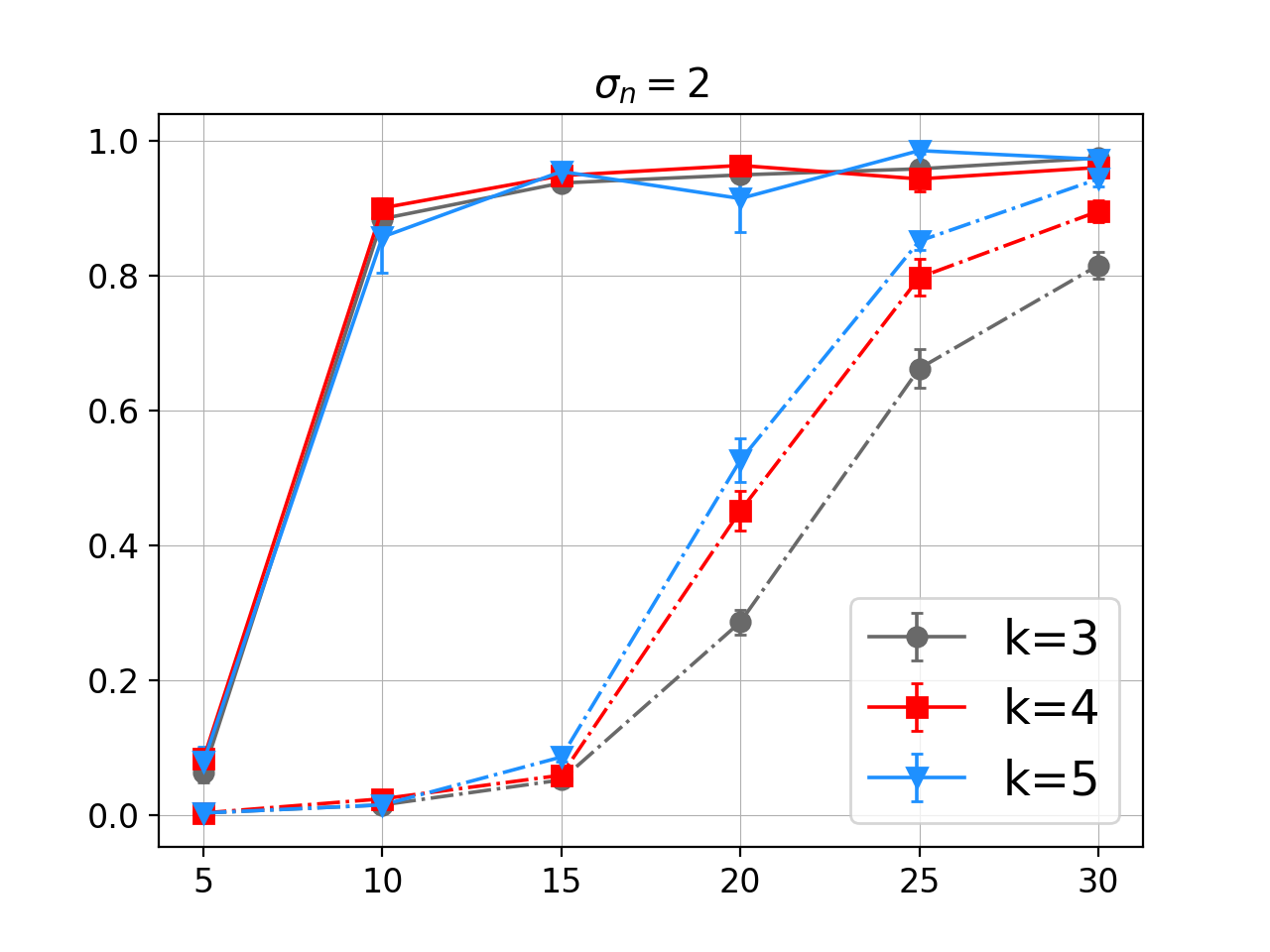}\includegraphics{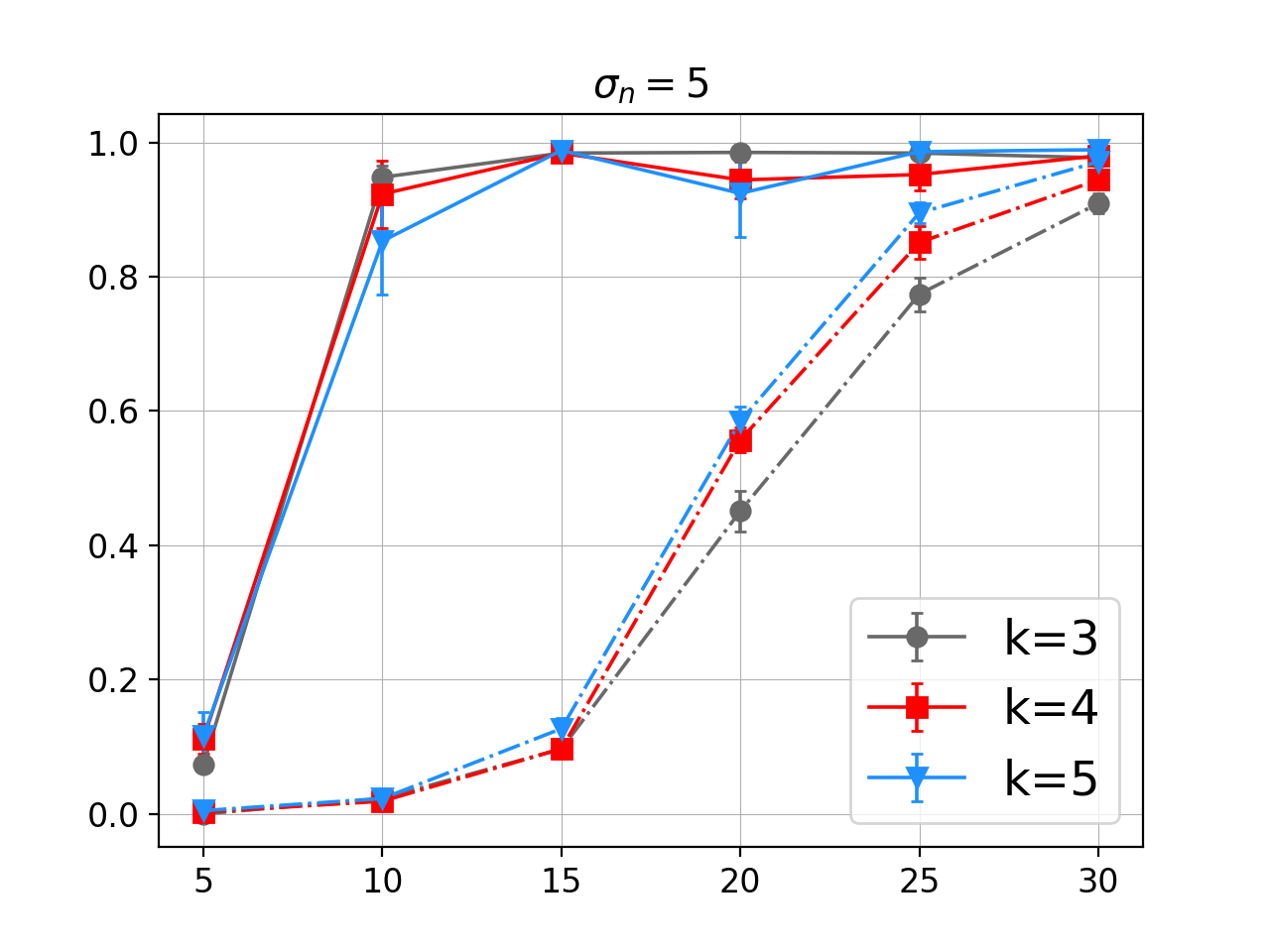}} 
		\caption{The graphs indicate the portion of systems stabilized by Algorithm~\ref{algo} versus the joint-stabilization length $\stablength$. Each plot corresponds to a value $\ditherscale \in \left\{ 0.3, 0.9, 2,5 \right\}$, for different number of epochs $\noofepochs=3,\textcolor{red}{4},\textcolor{blue}{5}$. Dashed curves present the stabilized portion by individual stabilization methods.}
		\label{fig:sigmaEta}
	\end{figure}
	
	\begin{figure}
		\centering
		\resizebox*{14.5cm}{!}{\scalebox{1.17}{\includegraphics{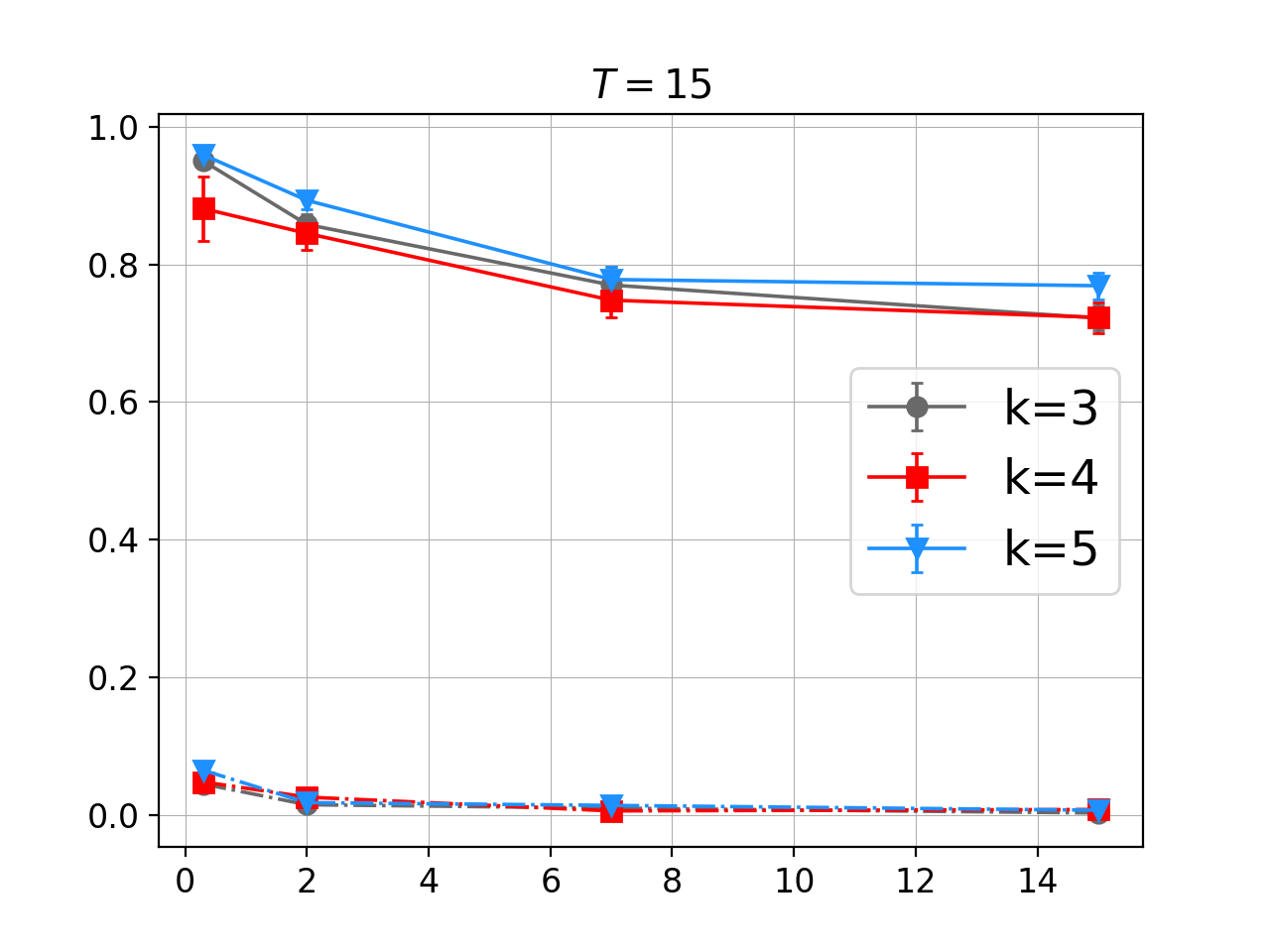}}\includegraphics{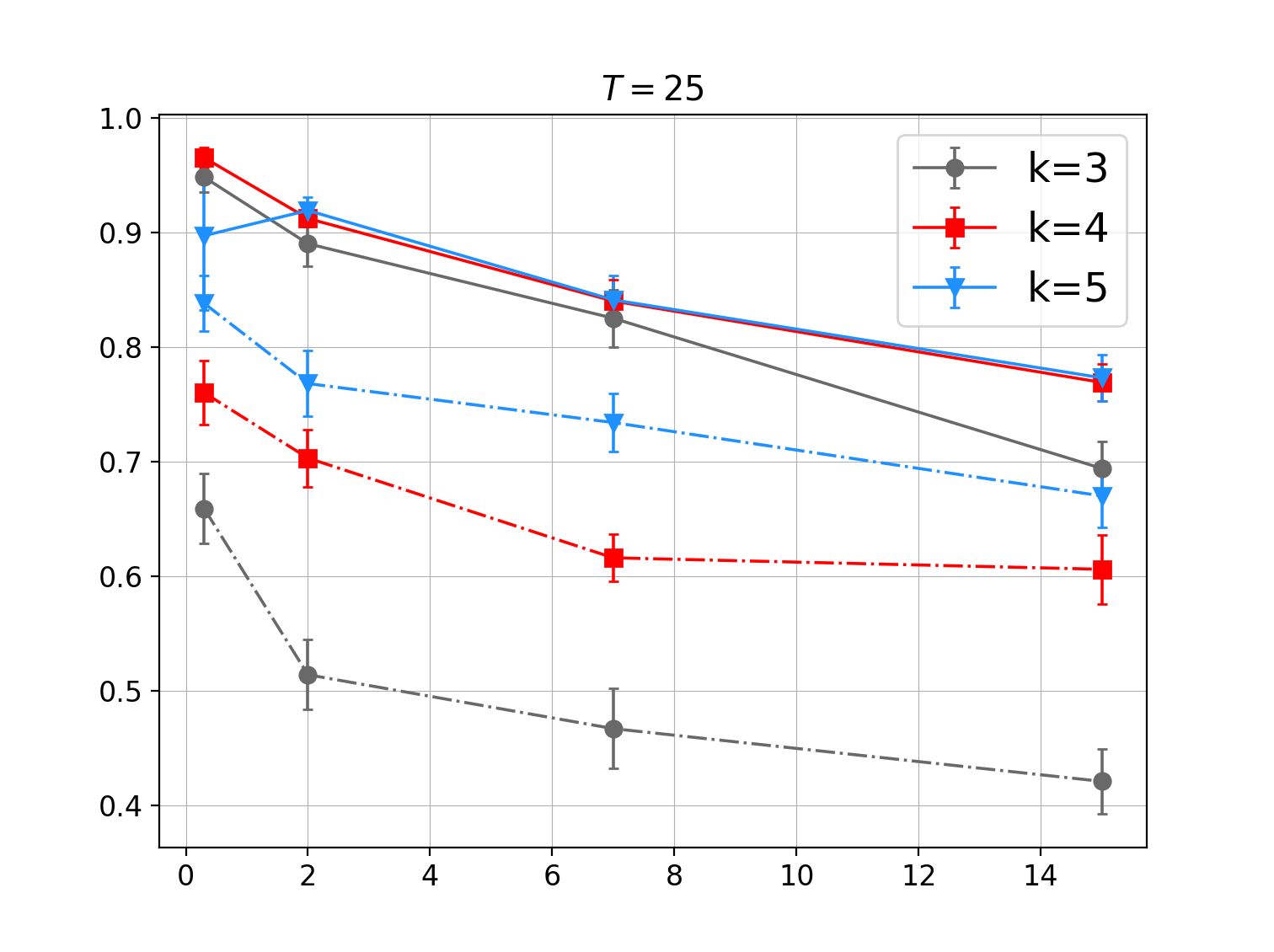}} 
		\caption{The fraction of stabilized dynamical systems by Algorithm~\ref{algo} is plotted versus $r$, where the matrices in the Riccati equation are $\Qmatrix=I_{\statedim}, \Rmatrix = r I_{\controldim}$. The graphs correspond to the joint-stabilization lengths $\stablength=15,25$, for different number of epochs $\noofepochs=3,\textcolor{red}{4},\textcolor{blue}{5}$. As before, dashed lines depict the fraction of individually stabilized systems under the same conditions. Note that the ranges of y-axes are different for the graphs.}
		\label{fig:QR}
	\end{figure}

	\section{Analysis of Algorithm~\ref{algo}} \label{NumericalSection}
	
	In this section, we report various analyses for studying Algorithm~\ref{algo} and its performance for successfully learning to stabilize all the systems. In the sequel, effects of different parameters in the success rates of the algorithm for utilizing similarities of dynamical systems are studied. 
	
	We simulate $\nooftasks=100$ different linear dynamical systems of dimensions $\statedim=10$ and $\controldim=6$. Entries of the $\noofbases=5$ basis matrices $\truebasismat{j}$, as well as the coefficients $\trueweights{i}{j}$ in \eqref{SharedBasesAssumpEq}, are generated randomly. Then, the true dynamics parameters of all systems $\truth{i}, i =1,\cdots, \nooftasks$, are calculated according to \eqref{SharedBasesAssumpEq}. In the above procedure, the open-loop spectral radius values $\left\{\eigmax{\trueA{i}}\right\}_{i =1}^\nooftasks$ vary in the range from $1.2$ to $1.5$. 
	
	Therefore, all systems are open-loop unstable and proper design of stabilizing policies are required to steer those eigenvalues of $\trueA{i}$ outside the unit-circle towards inside of that, through the input matrices $\trueB{i}, i =1,\cdots, \nooftasks$. Furthermore, all stochastic disturbance vectors $\noise{i}{t}$ are independent and have the distribution $\normaldist{0}{\sigma_\xi^2 I_{\statedim}}$, for $\sigma_\xi=2$. Unless otherwise explicitly mentioned, the matrices $\Qmatrix,\Rmatrix$ used in the Riccati equation~\eqref{RiccatiEq} are $\Qmatrix=I_{\statedim}$ and $\Rmatrix=r I_{\controldim}$, where $r=0.25$. In all experiments, for each curve, the numbers are generated by running the simulation with $10$ different random seeds and the standard errors bars are plotted. 
	
	In order to solve the optimization problem in \eqref{JointLearningStepEq}, Adam optimizer in PyTorch is employed~\cite{kingma2015adam}. The gradient steps are properly scaled according to a mini-batch stochastic gradient descent approach. For solving the discrete-time Ricatti equation in \eqref{RiccatiEq}, we use 
	the SciPy library in Python. On the other hand, to compare to individual stabilization of dynamical systems, the least-squares method that is known to be effective, is applied~\cite{faradonbeh2018bfinite,faradonbeh2019randomized}. 
	
	In Fig.~\ref{fig:sigmaG}, the fraction of $\nooftasks=100$ systems that Algorithm~\ref{algo} successfully stabilizes is reported versus the joint-stabilization length $\stablength$. Each graph corresponds to different value of $\FBscale$, while the scaling of the dither signal is fixed to $\ditherscale=2$. Each curve represents a fixed number of epochs $\noofepochs \in \left\{ 3,4,5 \right\}$, as specified in the legends. Moreover, the dashed curves depict corresponding stabilization rates through individual learning methods that apply to each system separately. These plots indicate that small values of $\FBscale$ leads to effective joint-stabilization, remarkably faster than individual stabilization of the systems. 
	
	However, for large $\FBscale$ , the random feedback matrices $\Gainmatrix{j}$ diversify the systems and deteriorate their similarities. Further, the eigen-spectrum of the resulting closed-loop matrices are so different that the magnitudes of the state vectors of $\nooftasks$ systems scale differently. Therefore, such chaotic dynamical behaviors defeat the purpose of joint-learning and deteriorate the performance of the joint stabilization algorithm. This phenomena indicates that in practice the value of $\FBscale$ needs to be selected carefully and sufficiently small. Still, it cannot be too small, since random feedback matrices of negligible magnitude practically reduce the number of epochs $\noofepochs$ to $1$, which does not provide an acceptable performance.
	
	Next, the performance of Algorithm~\ref{algo} for different values of $\ditherscale$ are depicted in Fig.~\ref{fig:sigmaEta}. The scale of random feedback matrices $\FBscale=0.3$ is fixed, while for $\ditherscale$ four values of $0.3, 0.9, 2 , 5$ are performed, each of which in one graph. Again, different number of epochs $\noofepochs$ are distinguished by colors. Similar to the previous figure, the fraction of dynamical systems that are stabilized by individual learning methods are reflected by dashed curves in each plot. According to Fig.~\ref{fig:sigmaEta}, larger $\ditherscale$ values help the joint-stabilization algorithm to success slightly faster. 
	
	The above positive effect of $\ditherscale$ on the success rate of Algorithm~\ref{algo} is reasonable because larger dither signals $\dither{i}{t}$ increase explorations by randomizing $\control{i}{t}$ around the state-feedback vectors of $\Gainmatrix{j}\state{i}{t}$. However, this effect is not very large since instabilities during the implementation of the algorithms makes exponentially growing state vectors dominant, compared to $\dither{i}{t}$. Furthermore, since the stochastic dither signals are drawn independently at every time step and so does not have any significant future consequences, $\ditherscale $ can be relatively large without deteriorating influences on the performance. Note that this is totally opposite to the effect of large $\FBscale$ we discussed in the previous paragraph.
	
	Fig.~\ref{fig:QR} contains the rates of successful stabilization among all $\nooftasks=100$ linear dynamical systems. The horizontal axes reflect different values of $r$, which is the scaling of the matrix $\Rmatrix=rI_{\controldim}$ that is used in~\eqref{RiccatiEq} for calculating $\RiccatiSolution{\indivestimate{i}{\stablength}}$ and $\OptGain{\indivestimate{i}{\stablength}}$. As can be seen in Fig.~\ref{fig:QR}, smaller values of $r$ yield to better stabilization rates. That is intuitively consistent with the roles of $\Qmatrix,\Rmatrix$ in \eqref{RiccatiEq} for linear dynamical systems with quadratic cost functions~\cite{kailath2000linear}. Indeed, smaller $\Rmatrix$ matrices penalize the control actions $\control{i}{t}$ less severely compared to the penalty for state vectors through $\Qmatrix=I_{\statedim}$. Accordingly,  for smaller $r$, the control inputs have more freedom for stabilizing the dynamical systems.
	
	\section{Concluding Remarks and Future Work} 
	
	We studied the problem of stabilization of multiple linear systems with unknown dynamics matrices. The focus is on utilization of similarities of the dynamical systems for joint learning-based stabilization of all of them. Importantly, the framework consists of a general setting that the dynamics matrices are \emph{unknown} linear functions of some \emph{unknown} matrices. We propose a joint-stabilization algorithm for learning to stabilize all the systems, despite the compound uncertainties about the dynamics matrices. The performance of the algorithm is studied in extensive numerical experiments, showing that the main objective of \emph{fast} stabilization is achieved significantly better compared to individual learning-based methods.
	
	To the authors' knowledge, this work is the first that considers joint learning-based control of multiple unknown systems in a general setting. Accordingly, the roadmap towards several interesting problems are pointed out in this paper. An immediate future work of interest is that of theoretical analysis of the algorithm and its provable effectiveness. Extending the framework to other problems in multi-agent control such as cooperative and non-cooperative distributed adaptive control of dynamical systems is another direction for future studies.

\end{document}